\documentclass[sigconf,natbib=false,screen,authorversion,nonacm,review=false,timestamp=false]{acmart}

\usepackage{sigir26-topic-adapters-as-judges}
\graphicspath{{sigir26-topic-adapters-as-judges-figures}}

\usepackage{enumitem}
\usepackage{layouts}
\begin{document}
\title[Topic-Specific Classifiers are Better Relevance~Judges than Prompted~LLMs]{Topic-Specific Classifiers are Better \\ Relevance~Judges than Prompted~LLMs}

\author[L.\ Gienapp]{Lukas Gienapp}
\affiliation{
    \institution{University of Kassel, hessian.AI \& ScaDS.AI}
    \city{Kassel}
    \country{Germany}
}

\author[M.\ Potthast]{Martin Potthast}
\affiliation{
    \institution{University of Kassel, hessian.AI \& ScaDS.AI}
    \city{Kassel}
    \country{Germany}
}

\author[A. Yates]{Andrew Yates}
\affiliation{
    \institution{HLTCOE, Johns Hopkins University}
    \city{Baltimore}
    \country{USA}
}

\author[H.\ Scells]{Harrisen Scells}
\affiliation{
    \institution{University of Tübingen}
    \city{Tübingen}
    \country{Germany}
}

\author[E.\ Yang]{Eugene Yang}
\affiliation{
    \institution{HLTCOE, Johns Hopkins University}
    \city{Baltimore}
    \country{USA}
}

\begin{abstract}
The unjudged document problem, where systems that did not contribute to the original judgement pool may retrieve documents without a relevance judgement, is a key obstacle to the reuseability of test collections in information retrieval. While the de facto standard to deal with the problem is to treat unjudged documents as non-relevant, many alternatives have been proposed, such as the use of large language models~(LLMs) as a relevance judge (LLM-as-a-judge). However, this has been criticized, among other things, as circular, since the same LLM can be used as the ranker and the judge. We propose to train topic-specific relevance classifiers instead: By finetuning monoT5 with independent LoRA weight adaptation on the judgments of a single assessor for a single topic's pool, we align it to that assessor's notion of relevance for the topic. The system rankings obtained through our classifier's relevance judgments achieve a Spearmans'~$\rho$ correlation of~$>0.94$ with ground truth system rankings. As little as 128~initial human judgments per topic suffice to improve the comparability of models, compared to treating unjudged documents as non-relevant, while achieving more reliability than existing LLM-as-a-judge approaches. Topic-specific relevance classifiers are thus a lightweight and straightforward way to tackle the unjudged document problem, while maintaining human judgments as the gold standard for retrieval evaluation. Code, models, and data are made openly available.\footnote{\url{https://github.com/webis-de/conf26-topic-adapters-as-judges}}
\end{abstract}

%
%

\maketitle            

\section{Introduction}

An ad hoc retrieval experiment usually consists of a large collection of documents, several dozen search topics each specifying an information need, and relevance judgments on a representative sample of the documents for each topic~\cite{cleverdon1967cranfield,voorhees2018building}. For large collections, the manual effort involved in collecting relevance judgments usually prevents a complete assessment of all documents for all topics. Furthermore, most documents in a collection are typically not relevant to a specific topic. To find a compromise between evaluation costs and experimental validity, it is therefore necessary to build a focused subset of the documents for each topic, obtained by combining the top-$k$ documents from the various search engines to be evaluated, i.e., pooling, where $k$~can range from a handful to a few hundred documents per search engine. 

\begin{figure}[t]
    \centering
    \includegraphics[width=\linewidth]{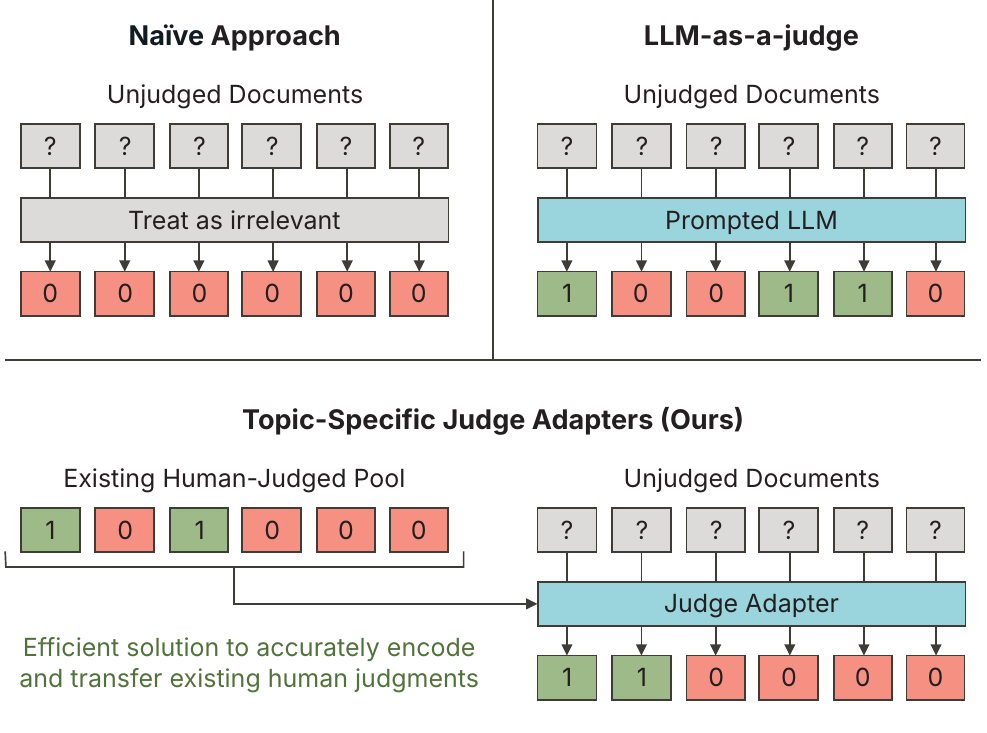}
    \caption{Conceptual illustration of how the na\"ive basline, the LLM-as-a-judge paradigm, and our proposed use of topic-specific judge adapters each operationalize the unjudged document problem.}
    \label{fig:placeholder}
\end{figure}

While this evaluation setup allows for a valid and fair comparison between the systems that initially contribute to the document pool of each topic, any new systems that are implemented after the experiment is complete are at a disadvantage~\cite{lipani2016fairness,parry2025variations}. These often retrieve documents that are not yet included in any given topic's pool, leading to the so-called unjudged document problem, namely the still mostly unresolved question of how to deal with such documents. Heuristic solutions such as simply omitting these documents, or deeming them non-relevant, have been proposed and frequently applied. However, none of the heuristics put new and initial systems on an equal footing. As research progresses and more unjudged but potentially relevant documents are uncovered by new retrieval systems, past evaluation experiments and test collections lose the capability to reliably distinguish between them, diminishing their reusability. At the same time, updating a test collection with new judgments is costly and the new judgments for a topic may be incongruent with those from the previous judges~\cite{voorhees2022can}.

Recent work has proposed to simply prompt instruction-tuned large language models~(LLMs) to supplement or completely replace human judgments~\cite{abbasiantaeb2025improving,upadhyay2025large,arabzadeh:2025,rahmani2025judging,abbasiantaeb2024can,upadhyay2024umbrela}, collectively called LLM-as-a-judge approaches.
However, any LLM used as a judge for evaluation (together with its judgment prompt) can simultaneously be used as a ranker in a retrieval system~\cite{soboroff2025dont,dietz2025principles}. This may lead to an LLM confirming its own ranking decisions~\cite{panickssery2024llm,wataoka2024self} during relevance judgment, biasing evaluation results~\cite{ye2025justice} in favor of a particular~LLM. Moreover, since LLMs inevitably encode social biases picked up from their pre-training data, evaluating retrieval models with LLMs may amplify them in retrieval models~\cite{dai2024bias}. 

We show that the key to address both, the unjudged document problem, and the bias due to using LLMs for both ranking and evaluation, is to train one relevance classifier for each topic based on the judgments of a single human assessor. A topic-specific classifier can only be used to judge documents for that topic, and it ideally closely approximates the notion of relevance originally applied by the human assessor. To demonstrate the effectiveness of this approach, we conduct experiments using the TREC Robust04~\cite{voorhees2004overview}, the TREC Deep Learning~2019~\cite{craswell2019overview}, and the TREC Deep Learning~2020~\cite{craswell2020overview} passage ranking test collections. We show that the system rankings produced by our topic-specific relevance classifiers strongly correlates with the rankings produced by the available human judgments. We analyze our approach with respect to shallow vs.\ deep human judgments, and in comparison to zero and few-shot LLM-as-a-judge approaches. Our relevance classifier correlates reliably with human judgments even under limited judgment depth, while outperforming LLM-as-a-judge approaches. These results strongly indicate the feasibility of using topic-specific relevance classifiers to automate retrieval evaluation after bootstrapping one classifier for every topic from gold standard human judgments.

\section{Related Work}
\label{sec:related-work}

We review the literature, relating our work to attempts at automating relevance judgments, with a particular focus on recent approaches, and how they differ from ours.

\paragraph{Ranking as Relevance Classification}
The notion of treating document ranking as a classification problem has been established early in information retrieval with the binary independence model~\cite{robertson:1976}. Despite the fact that this paradigm to ranking was, for a long time, less effective than other approaches, the idea of exploiting relevance judgments to tune relevance classifiers has also been proposed a long time ago, using logistic regression~\cite{gey:1994}, investigating active learning as a means to minimize the number of human judgments~\cite{lewis:1994}, and more recently, using modern re-ranking models~\cite{meng2026rerankers}. Unlike our approach, this notion of ranking attempts to train a {\em generic} relevance classifier for ranking instead of judging, which requires a lot of judgments, since the notion of relevance may differ wildly between topics and between human assessors. Later learning to rank approaches succeeded only once large query logs have been identified as source of implicit relevance feedback~\cite{joachims:2002}.

Topic-specific relevance classifiers for ranking have been studied at the TREC Filtering tracks~1994 to~2002~\cite{robertson:2002}, where participants trained many different kinds of classifiers to filter (classify) documents by their relevance with respect to various different settings.
Since each TREC topic is typically assigned to a single assessor, this implicitly means that all these classifiers also approximated the notion of relevance of that specific assessor, just like we propose in this paper. 

\paragraph{Efficient Relevance Annotation}
Assisting human assessors with automatic means during relevance annotation to increase the efficiency of judgment efforts has been a long-standing problem in information retrieval~\cite{soboroff2001ranking}. 
Documents worthwhile to annotate can be identified before labeling commences, e.g., through metric-based heuristics~\cite{moffat2006strategic,carterette2006minimal} or LLM-assisted annotation~\cite{takehi2025llm}. Or, akin to active learning, during labeling in iterative rounds~\cite{soboroff2021overview}, e.g., by enabling eDiscovery through technology-assisted review~\cite{cormack:2014}, at the TREC Total Recall tracks~2015 and~2016~\cite{grossman:2016}, and by training classifiers to judge documents outside domain-specific tasks~\citep{zhang2022survey}.
We envision the approach described in this paper not as an alternative to such efficient annotation techniques, but complementary to. Our approach allows for more reliable evaluation of test collections after the judgments have been collected, with many ways the initial judgments could be collected.

\paragraph{LLM-as-a-Judge}
LLMs have been demonstrated numerous times to have a general sense of relevance~\cite{pradeep2023rankzephyr,pradeep2023rankvicuna}. Many studies have thus investigated their applicability at collecting relevance judgments~\cite{abbasiantaeb2025improving,upadhyay2025large,rahmani2025judgeblender,rahmani2025judging,abbasiantaeb2024can,upadhyay2024umbrela} or even generating partial or fully synthetic test collections~\cite{bonifacio2022inpars,rahmani2025syndl,rahmani2024synthetic}. The general technique is to develop a prompt that encodes assessment instructions and to input this prompt alongside a query and unjudged documents. We argue that instead, topic specialization (i.e., learning from as much topic expertise as can be obtained) is key to automating relevance judgment in Cranfield-style evaluations, which the current zero-shot approach of LLM-as-a-judge forgoes in favor of generic instructions. While these studies have shown LLM judges to be correlated with human assessment, there are three main issues that current approaches do not address, namely that an LLM-as-a-judge
\Ni
exhibits biases and reliability issues~\cite{soboroff2025dont,panickssery2024llm,wataoka2024self,ye2025justice},
\Nii
is highly susceptible to prompt variations~\cite{arabzadeh:2025}, and
\Niii
can lead to invalid conclusions about evaluation, especially due to circularity~\cite{dietz2025principles,soboroff2025dont,}. Our approach is similar to the idea of LLMs-as-a-judge in that the judgment instructions are encoded into the model. However, rather than the prompt being used as the medium for storing what should be relevant, we directly encode what should be relevant into the parameters of the judging model. In doing so, we avoid the common pitfalls associated with LLMs-as-a-judge.

\paragraph{Relevance Judge Classifiers.}
The idea to predict relevance judgments and expand the overall set of relevance judgments was first proposed by~\citet{buettcher2007reliable}. The general idea was to train a binary SVM classifier using TF--IDF features from documents to determine relevance. There are three main issues with these experiments that we address in this paper:
\Ni
Their approach did not work well for deep assessments. Our approach effectively predicts relevance at much deeper depth.
\Nii
Their approach relies heavily on lexical features. Today, effective ranking models use much more complex, semantic scoring functions and our approach models semantics much more effectively.
\Niii 
Their approach trains classifiers from scratch, while we opt for adaptation of existing ranking models, leveraging their strong relevance priors.
\section{Topic-Specific Relevance Classifiers for Judging}\label{sec:methodological-approach}

Given a document corpus $D$ and a topic $t$ with query $q_t$ and partial binary relevance judgments $J_t = \{(q_t, d_i \in D, \{0,1\})\}_{i=1}^n, n \ll |D|$, our goal is to learn a topic-specific classifier $f_t: (q_t, d) \rightarrow [0,1]$ that assigns relevance scores to unjudged documents not in $J_t$, in the same way a human judge would when initially creating the test collection. Unlike traditional ranking models where cross-topic generalization is desirable, we deliberately overfit to the judgment patterns of a single assessor on a single topic. This inverts the standard bias-variance tradeoff in ranking models, and gives rise to judge models, where topic-specific accuracy matters more than generalization across the topic space.

We leverage pretrained ranking models as initialization rather than training classifiers from scratch. Models pretrained on large-scale relevance data provide strong semantic representations and implicit priors over document-query relevance~\cite{pradeep2023rankzephyr,pradeep2023rankvicuna}. We experiment with base ranking models from three archetypes: bi-encoders (representation-based), cross-encoders (interaction-based), and mono-decoders (generative scoring). Yet, full fine-tuning per topic is computationally prohibitive and produces large model weights unfit for distribution. We thus rely on LoRA~\cite{hu2022lora} to inject topic-specific behavior while keeping base model weights frozen. LoRA learns a low-rank approximation of the weight modifications attained during training. This allows each topic to receive independent adapter weights with no parameter sharing, preventing cross-topic interference. It further allows for lightweight distribution: one base model, many shareable topic-specific adapters.

\section{Experimental Setup}\label{sec:experimental-setup}

We conduct three experiments. The first investigates if, given a deep pool of judged data, training judge models is feasible. The second investigates how this changes if the pool changes to a shallow one, and how the predicted relevance judgments produced by a finetuned LoRA adapter (``judge adapter'') impact ranking evaluation results. Finally, the third investigates how the approach compares to LLM-as-a-judge methods, and if those can likewise benefit from topic-specific alignment through few-shot prompting.

\paragraph{Experiment 1: Model Judges from Deep Relevance Judgments}

We select 47 topics with at least 100 relevant documents from the deeply judged official \emph{eval} set of the TREC Robust04 dataset~\cite{voorhees2004overview}. Annotated graded relevance labels are cast to binary relevance. This yields, on average, \num{1350.31(444.23)} non-relevant and \num{183.02(78.76)} relevant documents per topic. We divide the available judgments in each topic into a train and a test set with a 80/20 split, with stratified sampling per label to ensure similar relevancy distribution in both sets. Then, for each of the three pointwise model archetypes bi-encoder~\cite{reimers2019sentence}, cross-encoder~\cite{reimers2019sentence}, and mono-decoder~\cite{nogueira2020document}, we select representative pretrained models as listed in \Cref{tab:non-llm-models}. We separately finetune topic-specific judge models for each topic and archetype, over 10 epochs, with a batch size of 64 query/doc pairs of a combined sequence length of 512 tokens, and a learning rate of 1e\textsuperscript{-4}. We train two version per topic: a native finetuning of the full model weights, and a LoRA adapter~\cite{hu2022lora} with parameters $r=64$ and $\alpha = 128$. The number of trainable parameters in each setting and for each model is also listed in \Cref{tab:non-llm-models}. Given the high class imbalance in each topic, we use a weighted MSE loss, with weights 0.95 and 0.05 for relevant/non-relevant labels. This corresponds to the inverse average relevance distribution in judgments.

\begin{table}[t]
\centering
\setlength{\tabcolsep}{2pt}
\renewcommand{\arraystretch}{.8}
\caption{Overview of the models investigated as topic-specific judges in this paper.}
\label{tab:non-llm-models}
\sffamily
\begin{tabularx}{\linewidth}{@{}lXrr}
\toprule
\multirow{2}{*}{\textbf{Archetype}}  & \multirow{2}{*}{\textbf{Model}}      & \multicolumn{2}{c}{\textbf{Parameters}} \\
                                                                            \cmidrule{3-4}
                    &                                                       & \multicolumn{1}{c}{Native}       & \multicolumn{1}{c}{LoRA}      \\
\midrule
Bi-Encoder          & \href{hf.co/sentence-transformers/all-MiniLM-L6-v2}{all-MiniLM-L6-v2}     &  22.7\,M        &  5.4\,M        \\
Cross-Encoder       & \href{hf.co/cross-encoder/ms-marco-MiniLM-L12-v2}{ms-marco-MiniLM-L12-v2}  &  33.4\,M        &  6.1\,M        \\
Mono-Decoder        & \href{hf.co/castorini/monot5-base-msmarco-10k}{monot5-base-msmarco-10k}    & 222.0\,M        & 26.0\,M        \\
\bottomrule
\end{tabularx}
\end{table}

\begin{table}[t]
\centering
\setlength{\tabcolsep}{2pt}
\renewcommand{\arraystretch}{.8}
\caption{Overview of the LLMs backbones investigated for LLM-as-a-judge in this paper.}
\label{tab:llm-models}
\sffamily
\begin{tabularx}{\linewidth}{@{}lXcc}
\toprule
\textbf{Model}      & \textbf{Checkpoint}                                                                   & \textbf{Param.}    & \textbf{Reas.} \\
\midrule
Ministral-3         & \href{hf.co/mistralai/Ministral-3-14B-Instruct-2512}{Ministral-3-14B-Instruct-2512}   & \phantom{1}14\,B   & \xmark         \\
Qwen-3 Next         & \href{hf.co/Qwen/Qwen3-Next-80B-A3B-Instruct}{Qwen3-Next-80B-A3B-Instruct}            & \phantom{1}80\,B   & \xmark         \\
GPT-OSS             & \href{hf.co/openai/gpt-oss-120b}{gpt-oss-120b}                                        &           120\,B   & \cmark         \\
MiniMax-M2.1        & \href{hf.co/MiniMaxAI/MiniMax-M2.1}{MiniMax-M2.1}                                     &           229\,B   & \cmark         \\
GPT-4o              & gpt-4o-2024-11-20                                                                     &            ---     & \cmark         \\
\bottomrule
\end{tabularx}
\end{table}

\paragraph{Experiment 2: Model Judges from Shallow Relevance Judgments}

For the second experiment, we simulate a shallow train set by stratified sampling of $k \in \{64, 128, 192, 256\}$ documents per topic, of which, on average, 12.5\% are relevant, depending on availability of relevant documents in the pool. We then finetune topic-specific judge LoRA adapters based on the mono decoder model, using the same hyperparameters as before. We conduct experiments both on the previously used topics of TREC Robust04, and additionally on 29 and 28 topics with more than 50 relevant documents from the TREC Deep Learning 2019 (DL19,~\cite{craswell2019overview}) and 2020 (DL20,~\cite{craswell2020overview}) passage ranking tracks. Topics with higher counts of relevance labels were chosen to ensure sufficient label variety in both train and test sets even under subsampling conditions.

\paragraph{Experiment 3: Comparison to LLM-as-a-judge}

To compare our proposed judge adapters to established LLM-as-a-judge approaches, we compute relevance labels for the complete TREC DL20 passage pool, using all available topics. We apply two established zero-shot prompts for relevance judgment: 
\Ni UMBRELA~\cite{upadhyay2024umbrela} (yielding graded relevance labels which we cast to binary) and
\Nii the best human-written prompt from \citet{arabzadeh:2025} (directly yielding binary relevance labels).
Furthermore, we test if access to topic-specific information yields improvements in LLM-as-a-judge by adding 8 few-shot examples (4 relevant, 4 non-relevant, in random order) to each prompt. Examples are drawn from the training set of our judge adapter for $t=256$ for comparability. We limit few-shot examples to 8 due to inference cost outweighing additional effectiveness benefit. Inference was carried out using varied open LLMs including both reasoning and non-reasoning models (see \Cref{tab:llm-models}). Additionally, we include the closed model GPT-4o for reference, since \citet{arabzadeh:2025} found it to be performing best for judging tasks.

\begin{figure*}[h!]
    \centering    
    \includegraphics[width=\linewidth]{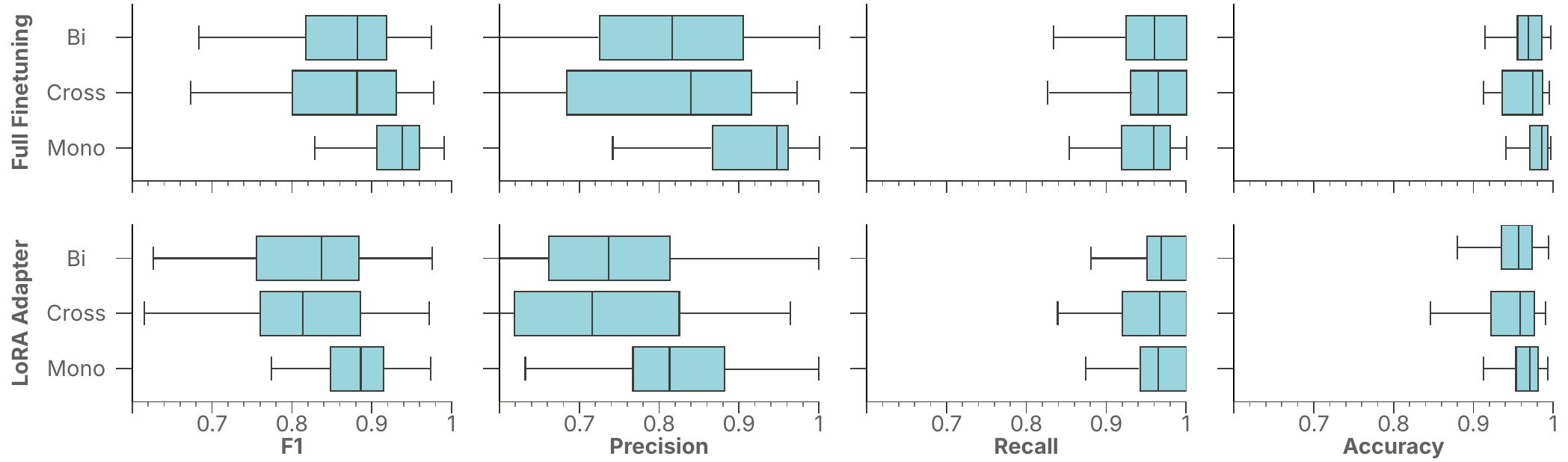}
    \caption{Effectiveness of our topic-specific ranker approach in terms of F$_1$, precision, recall, and accuracy when fine-tuning versus when using LoRA between different model architectures.}
    \label{fig:deep-pooling-metrics}
\end{figure*}

\section{Results \& Discussion}\label{sec:results-and-discussion}

We evaluate topic-specific relevance judges across three experimental conditions: 
\Ni 
deep pooling with abundant training data;
\Nii 
shallow pooling with limited annotations, and 
\Niii 
comparison against LLM-based approaches.

\subsection{Model Judges from Deep Relevance Judgments}

In the first experiment, we investigate whether ranking models can successfully be specialized as single-topic judges in a deeply judged setting, i.e., without restricting the quantity of training data.

\paragraph{Can ranking models be finetuned to judge topic-specific relevance?} Using the train/test set as described in \Cref{sec:experimental-setup}, we first evaluate the fully finetuned topic judge models w.r.t. precision, recall, F\textsubscript{1}, and accuracy. Results are shown in the first row of \Cref{fig:deep-pooling-metrics}. Overall, training topic-specific judge models is successful across all tested model types. Their effectiveness exceeds 0.88 average F\textsubscript{1} in all cases, with near-perfect recall and accuracy. Lower precision reflects class imbalance: only 6\% of Robust04 documents are relevant, causing false positive bias despite weighted loss training. Between the different model types, the mono-decoder was the most effective, exhibiting both the highest average scores and the lowest variance over topics. Cross-encoders are second, scoring slightly higher than bi-encoders on average, yet also exhibiting greater variance. 

\paragraph{How do LoRA adapters compare to full finetuning?}
To reduce the computational overhead in training judge models and make them ``portable'' for dissemination, we investigate the use of LoRA adapters instead of full finetuning. Using the same evaluation procedure as before, the second row of \Cref{fig:deep-pooling-metrics} shows that adapter-based judges only exhibit minimal deviation in evaluation metrics compared to their fully finetuned counterparts. Accuracy and recall remain the same, while a slight drop of 0.05 - 0.1 in precision (and thus F\textsubscript{1}) is observable, varying by model type. Cross-encoder adapters perform worse than bi-encoder and mono-decoder adapters, both in absolute terms and in relative difference to full finetuning. Mono-decoder adapters remain the best choice overall, and their adapter versions are a suitable replacement for full finetuning.

\subsection{Model Judges from Shallow Relevance Judgments}

We next investigate whether the effectiveness of judge adapters carries over to a shallow data regime. Using the best-performing ranker from before (monoT5), we tune judge adapters per topic on different datasets and amounts of training data.

\begin{figure*}[t]
    \centering
    \includegraphics[width=\linewidth]{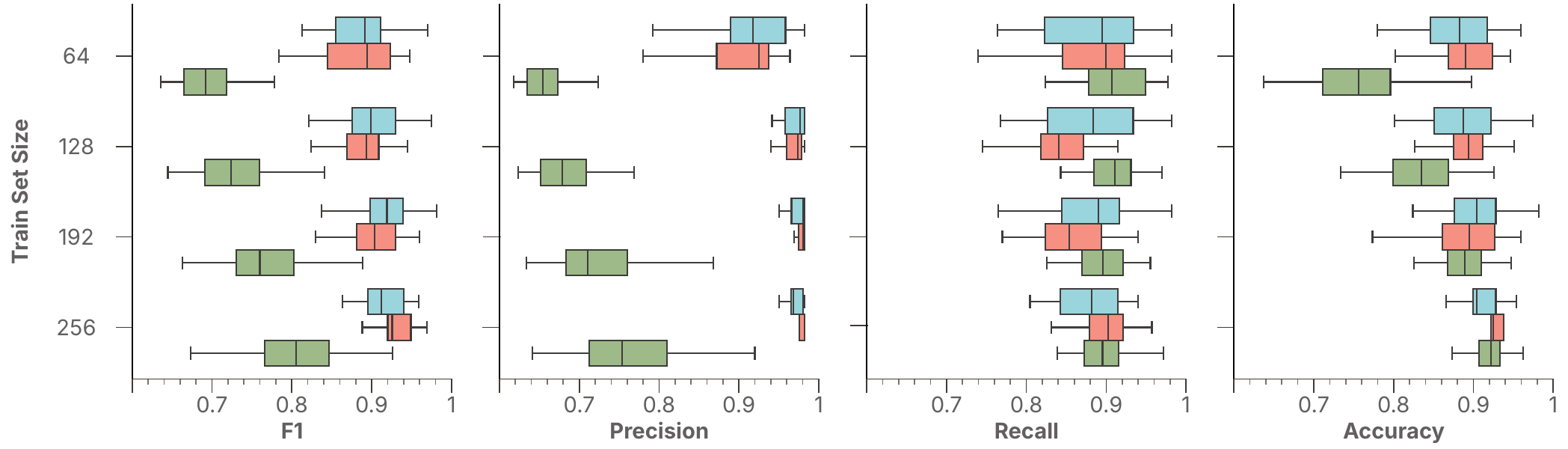}
    \caption{
        Precision, recall, F\textsubscript{1}, and Accuracy for binary relevance classification by adapter models using different amounts of training samples. Colors indicate datasets: \colordot{green-500} TREC Robust04, \colordot{blue-500} TREC DL19, \colordot{red-500} TREC DL20. The DL datasets are in-domain (MSMARCO) for the base ranking model, Robust04 is not.
    }
    \label{fig:shallow-pooling-metrics}
\end{figure*}

\paragraph{How much data is needed to tune a judge adapter?}
We compare the four variants of judge adapters, trained with 64, 128, 192, or 256 training examples, respectively, by their accuracy, precision, recall, and F\textsubscript{1} on the remaining judgment data per topic. \Cref{fig:shallow-pooling-metrics} shows that results are similar across all metrics and models for DL19 and DL20 datasets, reaching near-perfect precision at slightly lower but still excellent recall and accuracy. Recall on Robust04 remains comparable to the DL datasets, yet accuracy is reduced in low-data variants, and precision drops to between 0.7 and 0.75 on average. Overall, we see a trend of more training data yielding better predictions at diminishing returns for $>128$ samples. The high scores throughout assert the viability of judge adapters even on shallow pools.

\paragraph{Can judge adapters reliably fill missing relevance judgments?}
To measure the impact using judge adapters has on system rankings in a realistic scenario, we simulate missing judgments through reduced pools, and then compare the rank correlation between system rankings derived from predicted judgments to those derived from the ground-truth human relevance judgments. Reduced pools are simulated through run subsampling, where the top 100 documents of all chosen runs are assigned their ground-truth scores, i.e., the train set for the adapter model, while the additional documents from non-chosen runs are treated as `new', i.e., unjudged, and receive scores as predicted by the respective trained topic judge adapter. This allows us to simulate realistic unjudged documents patterns. In a bootstrapping-like setup, each such simulation is repeated 20 times with different seeds. Note that very small pools (1-3 runs) might not include enough documents to train some adapters with larger train set sizes, so these are omitted. As baseline and lower bound to compare to, we use evaluation results where missing relevance judgments are deemed irrelevant by default.

\begin{figure*}[ht!]
    \centering    
    \includegraphics[width=\linewidth]{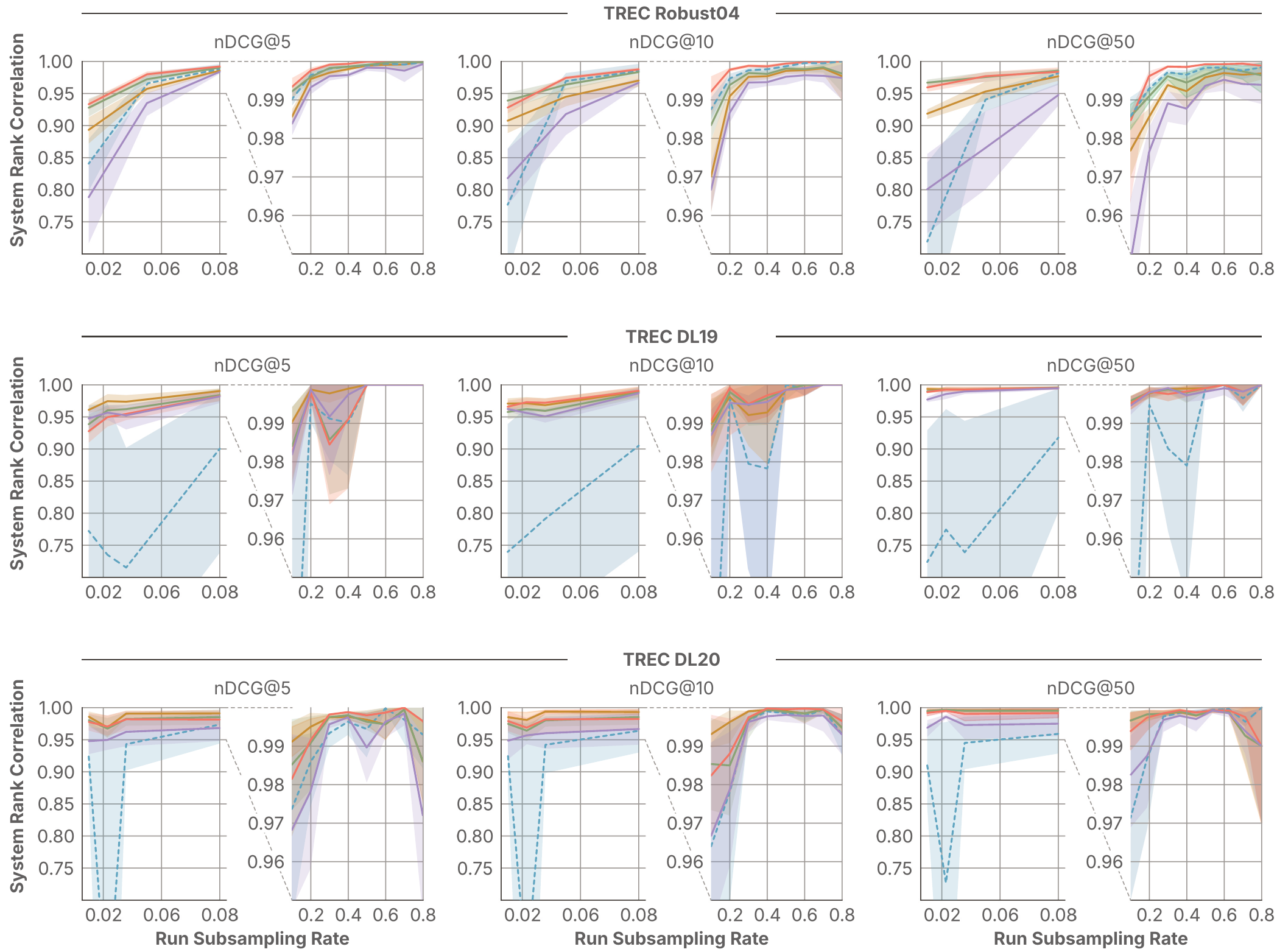}
    \caption{
        Spearmans' $\rho$ rank correlation of system orderings at different depths of nDCG to human judgments at different run-subsampling rates for simulated pooling. Missing relevance judgments infilled by
        \colordot{blue-500} assuming non-relevance (baseline, dashed), and judge adapter relevancy predictions for
        \colordot{red-500} $t=256$,
        \colordot{green-500} $t=192$,
        \colordot{yellow-500} $t=128$,
        \colordot{violet-500} $t=64$, where $t$ is adapter training set size. Bootstrapped uncertainty as shaded areas ($n=20$).
    }
    \label{fig:correlation-subsampling}
\end{figure*}

\Cref{fig:correlation-subsampling} plots the correlation values over run subsampling rates for the different datasets and adapter train set sizes. Throughout, the system rankings on predicted relevance judgments very strongly correlate with the ground truth system ranking. Here, too, higher amounts of training data yield more reliable judge adapters. As pooling sizes increases, i.e., the ratio of predicted to ground-truth relevance labels shifts towards the ground-truth, the ranking quickly approach perfect correlation ($>0.99$) at run subsampling rates of 0.2 and above across all datasets. However, also with extremely sparse pools of around 3-4 runs, given enough training data in each topic, judge adapters can successfully infer the correct system ranking, reaching correlation values of around 0.95. In comparison, the na\"ive common practice of deeming unjudged documents as non-relevant scores worse for extremely shallow pools, while difference even out as the proportion of ground-truth judgments increases. Judge adapters furthermore offer much stabler results, showing less variance across the 20 iterations of run subsampling. 

\paragraph{Does domain shift influence judge reliability?}

Comparing the effectiveness of judges across the three datasets as per \Cref{fig:shallow-pooling-metrics}, we can observe differences coinciding with data domain: judges are consistently better in the two in-domain evaluation datasets DL19 and DL20, which both derive from MSMARCO~\cite{nguyen2016msmarco}, the corpus the monoT5 base ranker was trained on, opposed to Robust04, which is more out-of-domain. The same trend is apparent in \Cref{fig:correlation-subsampling}, where both the absolute attained scores are lower for Robust04, and the relative difference to the baseline is smaller. Yet, while using effective in-domain base models is favorable for judge training, transferring rankers to other data domains is still feasible, but might require slightly more initial human judgments. 

\begin{table*}[t]
\centering
\caption{Comparison between original human judgments and and relevance predicted by each automatic approach (A = archetype, P = prompt, M = model, R = reasoning). Spearmans $\rho$ rank correlation between system rankings from original judgments and predicted relevance per topic, mean and 95\% CI over topics, for different depths $k$ of different retrieval measures; and Krippendorff's $\alpha$ agreement for nominal scales between original judgments and predicted relevance. Highest per archetype and column marked bold. Highest adapter colored {\color{green-700}green}; highest LLM-as-a-judge colored {\color{blue-700}blue}.}
\label{tab:llm-eval}
\renewcommand{\arraystretch}{1.2} 
\sffamily
\newcommand{\cg}{\cellcolor{green-200}}
\robustify\cg
\newcommand{\cb}{\cellcolor{blue-200}}
\robustify\cb

\setlength{\tabcolsep}{3pt}
\begin{tabularx}{\linewidth}{
    @{}l
    c@{\hspace{4pt}}
    X@{\hspace{4pt}}
    c@{\hspace{4pt}}
    *{2}{S[table-format=1.2(2), separate-uncertainty=true, retain-zero-uncertainty=true, detect-weight=true, detect-mode=true, detect-shape=true]}@{\hspace{8pt}}
    *{2}{S[table-format=1.2(2), separate-uncertainty=true, retain-zero-uncertainty=true, detect-weight=true, detect-mode=true, detect-shape=true]}@{\hspace{8pt}}
    *{2}{S[table-format=1.2(2), separate-uncertainty=true, retain-zero-uncertainty=true, detect-weight=true, detect-mode=true, detect-shape=true]}@{\hspace{8pt}}
    *{2}{S[table-format=1.2(2), separate-uncertainty=true, retain-zero-uncertainty=true, detect-weight=true, detect-mode=true, detect-shape=true]}@{\hspace{8pt}}
    S[table-format=-1.3,round-mode=places,round-precision=3, detect-weight=true, detect-mode=true, detect-shape=true]@{}
}
\toprule

\multicolumn{4}{c}{\bfseries Approach}                      
& \multicolumn{2}{c}{\bfseries $\rho$ / NDCG@k} 
& \multicolumn{2}{c}{\bfseries $\rho$ / Prec@k} 
& \multicolumn{2}{c}{\bfseries $\rho$ / Rec@k} 
& \multicolumn{2}{c}{\bfseries $\rho$ / RRank@k} 
& {\multirow[c]{2}{*}{$\alpha$}}
\\
\cmidrule(r){1-4}                                           \cmidrule(lr){5-6}              \cmidrule(lr){7-8}              \cmidrule(lr){9-10}             \cmidrule(lr){11-12}                                    
{A} & {P} & {M} & {R}                                       & {$k = 10$}    & {$k = 50$}    & {$k = 10$}    & {$k = 50$}    & { $k = 10$}   & {$k = 50$}    & {$k = 10$}    & {$k = 50$}    & {}            \\

\midrule
\multicolumn{4}{@{}l}{Baseline (0-filling)}                 &       0.33(17)   &       0.49(13)    &       0.52(10)  &       0.70(9)    &       0.52(10)   &       0.70(9)    &       0.52(19)   &       0.44(14)   &       -0.335331  \\
\midrule\multirow{10}{*}{\rotatebox[origin=c]{90}{\bfseries Zero-Shot LLM}}
& \cite{arabzadeh:2025}      & Ministral-3    & \xmark      &       0.76(7)    &       0.85(4)     &       0.84(4)   &       0.87(4)    &       0.84(4)    &       0.87(4)    &       0.76(8)    &       0.79(7)    &        0.485582  \\
& \cite{arabzadeh:2025}      & Qwen3-Next     & \xmark      &   \bf 0.81(5)    &   \bf 0.86(5)     &   \bf 0.86(4)   &   \bf 0.88(4)    &   \bf 0.86(4)    &   \bf 0.88(4)    &       0.82(7)    &       0.83(6)    &    \bf 0.546054  \\
& \cite{arabzadeh:2025}      & GPT-OSS        & \cmark      &       0.79(7)    &       0.84(5)     &       0.84(5)   &       0.88(5)    &       0.84(5)    &       0.88(5)    &   \bf 0.83(8)    &   \bf 0.84(7)    &        0.497943  \\
& \cite{arabzadeh:2025}      & Minimax-M1     & \cmark      &       0.77(7)    &       0.83(5)     &       0.84(5)   &       0.87(5)    &       0.84(5)    &       0.87(5)    &       0.78(9)    &       0.80(8)    &        0.485085  \\
& \cite{arabzadeh:2025}      & GPT-4o         & \cmark      &       0.73(8)    &       0.82(6)     &       0.79(6)   &       0.85(6)    &       0.79(6)    &       0.85(6)    &       0.81(8)    &       0.82(7)    &        0.395021  \\
\cmidrule(){2-13}
& \cite{upadhyay2024umbrela} & Ministral-3    & \xmark      &       0.82(5)    &       0.84(6)    &        0.87(3)   &       0.87(5)    &       0.87(3)    &       0.87(5)    &       0.81(7)    &       0.81(7)    &        0.332389  \\
& \cite{upadhyay2024umbrela} & Qwen3-Next     & \xmark      &       0.82(5)    &       0.85(5)    &        0.87(3)   &       0.88(4)    &       0.87(3)    &       0.88(4)    &       0.81(6)    &       0.82(6)    &        0.533297  \\
& \cite{upadhyay2024umbrela} & GPT-OSS        & \cmark      &       0.82(5)    &       0.85(6)    &        0.88(4)   &       0.88(5)    &       0.88(4)    &       0.88(5)    &       0.81(7)    &       0.81(7)    &        0.490345  \\
& \cite{upadhyay2024umbrela} & Minimax-M1     & \cmark      &   \bf 0.84(5)    &       0.84(7)    &    \bf 0.89(3)   &       0.88(5)    &   \bf 0.89(3)    &       0.88(5)    &       0.83(7)    &       0.84(6)    &        0.462812  \\
& \cite{upadhyay2024umbrela} & GPT-4o         & \cmark      &       0.83(5)    &   \bf 0.87(4)    &        0.88(3)   &\cb\bf 0.90(4)    &       0.88(3)    &\cb\bf 0.90(4)    &   \bf 0.83(6)    &   \bf 0.84(6)    &    \bf 0.538728  \\
\midrule\multirow{10}{*}{\rotatebox[origin=c]{90}{\bfseries Few-Shot LLM ($n = 8$)}}
& \cite{arabzadeh:2025}      & Ministral-3    & \xmark      &       0.64(8)    &       0.79(6)     &       0.75(6)   &       0.86(4)    &       0.75(6)    &       0.86(4)    &       0.60(10)   &       0.61(10)   &        0.319731  \\
& \cite{arabzadeh:2025}      & Qwen3-Next     & \xmark      &   \bf 0.81(6)    &   \bf 0.86(4)     &   \bf 0.87(4)   &   \bf 0.89(4)    &   \bf 0.87(4)    &   \bf 0.89(4)    &       0.82(8)    &       0.83(8)    & \cb\bf 0.575372  \\
& \cite{arabzadeh:2025}      & GPT-OSS        & \cmark      &       0.81(7)    &       0.86(5)     &       0.87(4)   &       0.89(4)    &       0.87(4)    &       0.89(4)    &\cb\bf 0.85(6)    &\cb\bf 0.86(6)    &        0.528008  \\
& \cite{arabzadeh:2025}      & Minimax-M1     & \cmark      &       0.79(7)    &       0.84(6)     &       0.85(5)   &       0.87(5)    &       0.85(5)    &       0.87(5)    &       0.80(9)    &       0.81(8)    &        0.514433  \\
& \cite{arabzadeh:2025}      & GPT-4o         & \cmark      &       0.73(9)    &       0.82(7)     &       0.79(7)   &       0.86(6)    &       0.79(7)    &       0.86(6)    &       0.81(9)    &       0.82(8)    &        0.425477  \\
\cmidrule(){2-13}
& \cite{upadhyay2024umbrela} & Ministral-3    & \xmark      &       0.81(5)    &       0.87(4)     &       0.87(4)   &       0.89(4)    &       0.87(4)    &       0.89(4)    &       0.81(7)    &       0.81(6)    &        0.471275  \\
& \cite{upadhyay2024umbrela} & Qwen3-Next     & \xmark      &\cb\bf 0.85(5)    &\cb\bf 0.87(4)     &\cb\bf 0.90(3)   &   \bf 0.89(4)    &\cb\bf 0.90(3)    &   \bf 0.89(4)    &   \bf 0.84(7)    &   \bf 0.85(6)    &    \bf 0.526543  \\
& \cite{upadhyay2024umbrela} & GPT-OSS        & \cmark      &       0.85(5)    &       0.83(8)     &       0.89(3)   &       0.87(6)    &       0.89(3)    &       0.87(6)    &       0.81(7)    &       0.81(7)    &        0.489223  \\
& \cite{upadhyay2024umbrela} & Minimax-M1     & \cmark      &       0.83(5)    &       0.86(6)     &       0.88(3)   &       0.88(4)    &       0.88(3)    &       0.88(4)    &       0.83(7)    &       0.83(6)    &        0.468209  \\
& \cite{upadhyay2024umbrela} & GPT-4o         & \cmark      &       0.83(5)    &       0.87(5)     &       0.88(4)   &       0.89(4)    &       0.88(4)    &       0.89(4)    &       0.81(8)    &       0.83(7)    &        0.507997  \\
\midrule\multirow{4}{*}{\rotatebox[origin=c]{90}{\textbf{Adapt.} (Ours)}}
& \multicolumn{3}{l}{monoT5 ($t = 64$)}                     &       0.87(5)    &       0.96(2)     &       0.91(3)   &       0.98(1)    &       0.91(3)    &       0.98(1)    &\cg\bf 0.92(4)    &\cg\bf 0.93(4)    &        0.809898  \\
& \multicolumn{3}{l}{monoT5 ($t = 128$)}                    &       0.84(5)    &       0.96(1)     &       0.89(4)   &       0.98(1)    &       0.89(4)    &       0.98(1)    &       0.86(6)    &       0.87(6)    &        0.807580  \\
& \multicolumn{3}{l}{monoT5 ($t = 192$)}                    &       0.85(5)    &       0.96(2)     &       0.90(4)   &       0.98(1)    &       0.90(4)    &       0.98(1)    &       0.89(6)    &       0.90(6)    &        0.830581  \\
& \multicolumn{3}{l}{monoT5 ($t = 256$)}                    &\cg\bf 0.89(4)    &\cg\bf 0.97(1)     &\cg\bf 0.93(3)   &\cg\bf 0.99(0)    &\cg\bf 0.93(3)    &\cg\bf 0.99(0)    &       0.91(6)    &       0.91(6)    & \cg\bf 0.876776  \\
\bottomrule
\end{tabularx}
\end{table*}

\subsection{Comparison to LLM-as-a-judge}

In the third experiment, we compare our proposed approach to established LLM-as-a-judge methods, which have recently been proposed as effective solutions to the unjudged document problem. Furthermore, given the demonstrated benefits of topic specialization for judges, we investigate whether establishing specific topic context through few-shot prompting increases LLM-as-a-judge effectiveness. As before, we sample 256 documents per topic as train set for all topics of DL20. These retain their ground truth relevance annotations, and are used both to train judge adapters, as well as to sample 4 relevant and 4 non-relevant few-shot examples from. We then use each of the five different LLMs, each with both prompt variants and in zero- and few-shot configurations, to judge the remaining unjudged documents from the TREC DL20 pool. 

To evaluate label source quality, we compare system rankings induced by different label sources at varying retrieval depths ($k \in \{5, 10, 50\}$) across five metrics: Reciprocal Rank (RR), normalized Discounted Cumulative Gain (NDCG), Mean Average Precision (MAP), Recall, and Precision. The systems being ranked are the original TREC DL20 submissions; only the relevance judgments differ across label sources. For each metric, we calculate Spearman's $\rho$ rank correlation between system rankings computed using complete human label ground truth versus system rankings computed using the hybrid judgment sets (initial 256 human judgments plus automatic labels for remaining documents). Correlation is measured per-topic and then averaged over all evaluation topics of TREC DL20. Additionally, we measure direct inter-annotator agreement between each automatic labeling approach and the ground truth labels using Krippendorff's $\alpha$ for nominal scales. Agreement is measured on the complete label set across all topics. \Cref{tab:llm-eval} lists results. We interpret our findings through the lens of LLM evaluation tropes proposed by~\citet{dietz2025principles}, which identify common failure modes when using LLMs as evaluators.

\paragraph{How do judge adapters compare to LLM-as-a-judge?}

Judge adapters provide the best approximation of human judgments, exhibiting both highest rank correlations and inter-annotator agreement ($\alpha = 0.81$--$0.88$) across all configurations. At $k=10$, adapters achieve $\rho \geq 0.84$ across all metrics; at $k=50$, correlations reach $\rho \geq 0.96$ for NDCG, Precision, and Recall. RR@k correlations are notably lower ($\rho = 0.87$--$0.93$). Across all metrics and evaluation depths, judge adapters consistently outscore LLM-as-a-judge approaches for completing the judgment pool. 

While differences in system rankings between judge adapters and LLM-as-a-judge remain moderate (\mbox{$\Delta_{k=10}\approx 0.04, \Delta_{k=50}\approx 0.09$}), the inter-annotator agreement for LLMs is much lower (\mbox{$\alpha = 0.32$--$0.58$}) compared to adapters. Generally, $\alpha \geq 0.80$ indicates good reliability, while $0.67 \leq \alpha < 0.80$ allows only tentative conclusions~\cite{artstein2008intercoder}. This exemplifies what \citet{dietz2025principles} term the \emph{Ignored Label Correlation} trope: high system-level correlation can obscure important disagreements at the judgment level~\cite{faggioli2023perspectives}. LLMs seemingly apply a consistent but more lenient relevance threshold, flagging substantially more documents as relevant than human annotators, but do so systematically across all systems rather than favoring particular retrieval approaches. Thus, LLMs preserve the relative ordering of systems. Examining correlation patterns across depths $k \in \{10, 50\}$, this further indicates that with deeper unjudged pools, LLMs may introduce more false positive ratings of relevance, thus producing lower system-level correlation compared to the adapter approaches. We further observe that RR@k exhibits the largest spread across label sources, particularly when few-shot prompting destabilizes smaller models (e.g., Ministral-3 drops to $\rho = 0.60$ for RR@10). NDCG@k, Recall@k, and Precision@k remain comparatively stable. This suggests that disagreements between label sources primarily manifest in the very top ranks, affecting which systems appear best at identifying the single most relevant document, while set-level assessment remains robust.

Furthermore, the naive baseline of assuming non-relevance, which exhibits negative agreement ($\alpha = -0.34$), achieves passable system-level correlation at $k=50$ (Precision, Recall: $\rho = 0.70$). This provides further evidence that system-level orderings are rather robust while on the label-level, judgments can be drastically different, corroborating the interpretation that $\alpha$-values are more indicative of label source quality.

\newpage
\paragraph{Does the backbone LLM model influence LLM-as-a-judge effectiveness?}
Model choice demonstrates measurable but inconsistent effects on both ranking correlation and inter-annotator agreement. No clear pattern emerges favoring 
larger or reasoning-capable models: Qwen3-Next, a smaller open-weight model without reasoning, obtains the highest $\alpha$ values ($0.53$--$0.58$), while GPT-4o, identified as best annotation model by \citet{arabzadeh:2025} shows surprisingly variable agreement ($\alpha = 0.40$--$0.54$) depending on prompt choice, even at its much higher parameter count and reasoning capability. This suggests that relevance assessment may be a calibration task rather than a reasoning task. Inter-annotator agreement ranges from $\alpha = 0.33$ to $0.55$ across zero-shot configurations. This pattern suggests that while different models disagree on the threshold for relevance, they maintain similar system ranking behavior, i.e., do not favor particular documents or contributing systems, which would manifest in diverging system orderings. 

However, this similarity may indicate susceptibility to the \emph{Loss of Variety of Opinion} trope: if all LLMs converge on similar ranking behaviors due to shared training paradigms or data sources, the community loses the diversity of perspectives that human assessors provide, and which topic-specific adapters can reproduce. Furthermore, while the training data for our adapter backbone model is known, we cannot guarantee that the open-weight LLMs did not see test set data from TREC-DL20 during pretraining, i.e., a threat of \emph{Test Set Leak} as identified by \citet{dietz2025principles}, inflating their scoring effectiveness on this collection. 

\paragraph{Does prompt variation influence LLM-as-a-judge effectiveness?}
Comparing the two prompt variants across identical model configurations, we observe minimal systematic differences in ranking correlations. Both the prompt of \citet{arabzadeh:2025} and \citet{upadhyay2024umbrela} achieve comparable $\rho$ values (typically within $0.01$--$0.03$), with neither consistently outperforming the other. This stability suggests that system ranking validity is relatively robust to prompt formulation. However, inter-annotator agreement varies substantially ($\Delta\alpha$ up to $0.15$ for identical models with different prompts, e.g., Ministral-3), indicating that prompt design affects the relevance threshold applied by the LLM. Comparing zero-shot and few-shot variants, few-shot prompting yields mixed results. For Qwen3-Next with \citeauthor{arabzadeh:2025}'s (\citeyear{arabzadeh:2025}) prompt, few-shot examples increase agreement from $\alpha = 0.55$ to $\alpha = 0.58$. However, for Ministral-3 with the same prompt, few-shot prompting causes substantial degradation (RR@10 drops from $0.76$ to $0.60$; $\alpha$ drops from $0.49$ to $0.32$). Overall, few-shot prompting at $n = 8$ examples does not yield consistent improvements. At the same time, it increases token budget and computational cost. The susceptibility of $\alpha$-values to model and prompt choice illustrates \citeauthor{dietz2025principles}'s (\citeyear{dietz2025principles}) \emph{LLM Evolution} trope, i.e., evaluation outcomes depend on model choice, configuration, and prompt usage.

\paragraph{Statistical Significance}
\Cref{tab:significance} lists significance levels for comparing adapters against the best-performing LLM per archetype-prompt configuration. These confirm that adapters significantly outperform LLM-as-a-judge at greater evaluation depths ($k = 50$, $p < 0.001$ in all cases), while with shallow depth ($k = 10$) differences are statistically significant for 3 out of the 4 comparisons for $t=256$, and 2 out of 4 for $t=64$. Since the comparisons are not independent (sharing the same system pool), and we test against the best-performing LLM rather than all configurations (a conservative choice), we do not apply multiple testing correction. However, even under Bonferroni correction, the $k = 50$ results remain highly significant ($0.05/16 = 0.003$). This reinforces that adapters are reliably advantageous to LLM-as-a-judge when evaluating beyond the very top ranks, i.e., extending (shallow) human-judged pools.

\begin{table}[t]
\centering
\caption{Significance level for comparing system correlations between adapter judges and best-performing LLM-as-a-judge approach per archetype and prompt for NDCG@10 and NDCG@50; paired one-sided $t$-test, $n=54$. Non-significant values ($p > 0.05$) greyed out.}
\label{tab:significance}
\small\sffamily
\newcommand{\cg}{\color{gray}}
\robustify\ns

\begin{tabularx}{\linewidth}{@{}l@{\hspace{4pt}}l@{\hspace{4pt}}X*{4}{S[table-format=1.4,round-mode=places,round-precision=4,detect-weight=true,detect-shape=true,detect-mode=true]}@{}}
\toprule
&    &                                                   & \multicolumn{4}{c@{}}{\bf Adapter (Ours)}           \\
                                                        \cmidrule(l){4-7}
&    &                                                   & {$t=64$} & {$t=128$}& {$t=192$}& {$t=256$}\\
\midrule
\multirow[c]{8}{*}{\rotatebox[origin=c]{90}{\bf LLM-as-a-judge}}
& \multirow[c]{4}{*}{\rotatebox[origin=c]{90}{NDCG@10}} 
      & Qwen3-Next, \cite{arabzadeh:2025}                 &     0.044021 & \cg 0.193785 & \cg 0.124780 &     0.015102 \\
&     & GPT-4o, \cite{upadhyay2024umbrela}                & \cg 0.107041 & \cg 0.322934 & \cg 0.215874 &     0.023833 \\
&     & Qwen3-Next, few-s., \cite{arabzadeh:2025}       &     0.049942 & \cg 0.210349 & \cg 0.140160 &     0.024902 \\
&     & Qwen3-Next, few-s., \cite{upadhyay2024umbrela}  & \cg 0.309613 & \cg 0.607447 & \cg 0.493067 & \cg 0.124205 \\
\cmidrule(l){2-7}
& \multirow[c]{4}{*}{\rotatebox[origin=c]{90}{NDCG@50}} 
      & Qwen3-Next, \cite{arabzadeh:2025}                 &     0.000012 &     0.000086 &     0.000053 &     0.000012 \\
&     & GPT-4o, \cite{upadhyay2024umbrela}                &     0.000045 &     0.000154 &     0.000106 &     0.000022 \\
&     & Qwen3-Next, few-s., \cite{arabzadeh:2025}       &     0.000002 &     0.000031 &     0.000017 &     0.000004 \\
&     & Qwen3-Next, few-s., \cite{upadhyay2024umbrela}  &     0.000076 &     0.000508 &     0.000384 &     0.000074 \\
\bottomrule
\end{tabularx}
\end{table}

\section{Practical Considerations for Using Judge Adapters}

To drive adoption of our approach within the wider research community, enabling the comparison of models on test collections that were previously not possible due to the retrieval of too many unjudged documents, we distribute code to train new judge adapters, and our existing judge adapters for DL19, DL20, and Robust04. However, we do so with the caveat that these models are only ever to be used for the specific use case of assessing unjudged documents for the intended topic. For practitioners looking to train judge adapters on new test collections, several considerations emerge from our experiments.

\paragraph{Data.} Our experiments suggest that between 100 and 200 judged documents per topic are needed to yield highly reliable judge adapters.
The natural class imbalance in judgment pools (typically between 3-12\% relevant documents) should be handled through adapting the loss function class weights to the specific test collection at hand. We found that oversampling strategies did not improve over loss weighting, and thus recommend it as the simpler approach.

\paragraph{Base Model.} The choice of base ranking model impacts both adapter effectiveness and computational requirements. Mono-deco\-ders consistently outperformed other architectures in our experiments, with a justifiable increase in parameter budget. LoRA-adapted monoT5 models can be trained and run for inference on most consumer GPUs, or even CPUs, in contrast to LLM-as-a-judge methods. Domain alignment between the base model's pretraining corpus and the target test collection improves effectiveness, though cross-domain transfer remains feasible with slightly more training data to compensate for domain shift.

\paragraph{Advantages.} 
\citet{dietz2025principles} identify fourteen evaluation tropes that can compromise the validity of LLM-based evaluation. Judge adapters inherently avoid several of these failure modes:
\begin{enumerate}[label=(\roman*),leftmargin=0pt, align=left,labelindent=\parindent, listparindent=\parindent, labelwidth=0pt, itemindent=!]
\setlength{\itemsep}{1ex}
    \item \textbf{Circularity avoidance:} Unlike LLM-as-a-judge approaches, where the same model can serve as both ranker and evaluator~\citep[Tropes \#1--2]{dietz2025principles}, judge adapters are topic-specific classifiers that cannot function as general-purpose rankers. Their deliberate overfitting to individual topics prevents dual use.
    \item \textbf{Preservation of assessor perspective:} LLM judges risk homogenizing relevance judgments and losing variety of opinion~\citep[Trope \#4]{dietz2025principles}. Judge adapters instead encode the specific relevance criteria of individual human assessors, preserving the diversity inherent in human judgment.
    \item \textbf{Reproducibility and stability:} LLM-based evaluators suffer from model evolution, where behavior changes across versions without notice~\citep[Trope \#7]{dietz2025principles}. Judge adapters produce deterministic, versioned outputs that remain stable over time, enabling reproducible evaluation.
    \item \textbf{Robustness to self-preference bias:} LLMs tend to favor text from their own model family~\citep[Trope \#3]{dietz2025principles}. Judge adapters are not subject to this bias, and instead closely reproduce human judgment.
    \item \textbf{Computational efficiency:} When scalability is required or inference costs are a concern, judge adapters offer substantial advantages, as LLM-as-a-judge approaches require significantly more computational resources, particularly with few-shot prompting. However, the cost of initially assembling a human-sourced label set has to be acknowledged as bottleneck.
\end{enumerate}

\paragraph{Limitations and Validity Constraints.} 
Judge adapters are designed specifically to extend existing judgment pools by approximating how a particular assessor would judge previously unseen documents for a specific topic. This design imposes strict validity constraints that must be observed to maintain evaluation integrity:

\begin{enumerate}[label=(\roman*),align=left,leftmargin=0pt, labelindent=\parindent, listparindent=\parindent, labelwidth=0pt, itemindent=!]
\setlength{\itemsep}{1ex}
    \item \textbf{No cross-topic transfer:} The assessment of similar topics or using judge adapters trained on one topic for another is invalid, violating the topic-specific alignment they are intended for.
    \item \textbf{No system integration:} The inclusion of judge adapters within a retrieval system, whether for re-ranking, filtering, or any purpose other than post-hoc evaluation, constitutes training on the test set. This directly instantiates the \emph{Circularity} trope~\citep[Trope \#1]{dietz2025principles} and invalidates any resulting evaluation.
    \item \textbf{No distillation or knowledge transfer:} Using judge adapter predictions to train or improve retrieval models would create the feedback loops characteristic of \emph{Goodhart-style Overfitting}~\citep[Trope \#10]{dietz2025principles}, where systems optimize for the evaluation signal rather than genuine relevance.
    \item \textbf{Test set leak considerations:} Unlike LLMs that may have been trained on publicly available test collections~\citep[Trope \#8]{dietz2025principles}, the base model for adapters (monoT5 in our case) has documented training data. However, practitioners training adapters on new collections should verify that their base model was not exposed to the target test collection.
    \item \textbf{Comparability to original judgments:} Expanding the judgment pool makes comparisons to systems that previously treated unjudged documents as non-relevant incompatible. Both systems must be evaluated on the new, expanded judgments.
\end{enumerate}

To support responsible use, we recommend releasing judge adapters alongside documentation specifying their intended scope. We encourage the community to treat these models as evaluation infrastructure rather than as components for system development, thereby preserving the separation between optimization and assessment that valid IR evaluation requires.

\paragraph{When to Prefer Alternatives.} 
Judge adapters are not universally optimal. LLM-as-a-judge approaches may be preferable when no existing human judgments are available for the target topic, precluding adapter training, or rapid prototyping across many new topics outweighs the need for precise alignment with specific assessor behavior. Practitioners should apply the guardrails recommended by \citet{dietz2025principles} to mitigate these associated validity risks.
\section{Conclusion}

The assumption that unjudged documents are non-relevant can lead to unreliable and incomparable results, especially as retrieval models improve beyond the systems that contributed to judgment pools. However, using LLM-as-a-judge approaches instead to extend the labeled pool yields poor agreement with human judgment.

We address this shortcoming through topic-specific judge adapters: LoRA-based fine-tuning of pretrained rankers that deliberately overfit to individual assessors' relevance criteria for individual topics. Aligning the predictions of pre-trained ranking models with topic-specific judgments addresses key limitations of previous LLM-based approaches, namely circularity, amplification of biases, and extreme resource usage. Judge adapters offer a reliable and scalable way to extend shallow pools to greater depth through automatic judgment.

Our experiments across three test collections demonstrate that judge adapters trained on as few as 64--256 examples per topic achieve strong correlation with ground truth system rankings, reaching $\rho \geq 0.96$ for NDCG, Precision, and Recall at $k = 50$, with more moderate but consistent advantages at $k = 10$ ($\rho \geq 0.84$). Paired significance tests confirm that adapters significantly outperform all LLM-as-a-judge configurations at higher evaluation depths ($p < 0.001$), while differences at $k = 10$ are in favor of adapter approaches but not always statistically significant. Judge adapters further exhibit substantially higher agreement with ground truth labels (Krippendorff's $\alpha = 0.81$--$0.88$ vs.\ $0.32$--$0.58$ for LLMs). Notably, reasoning-capable LLMs do not outperform simpler models on this task, suggesting relevance judgment is a calibration task rather than a reasoning task. While LLMs maintain moderately high system ranking correlations (up to $\rho = 0.90$ for the most effective configurations), they systematically over-label documents as relevant. Yet, they do so consistently across systems, preserving relative orderings.

These results establish judge adapters as a practical solution that preserves human judgment as the gold standard while extending test collection utility. More broadly, our findings reveal that topic-specific alignment, i.e., capturing 
how a particular assessor judges a particular topic, is both feasible with limited training data and more effective than seeking general-purpose relevance classifiers or relying on LLMs' pretrained relevance priors. Through topic-specific weight adaptation and the use of LoRA, topic adapters offer lightweight distribution and deterministic evaluation at negligible training and inference cost, while avoiding the evaluation tropes~\cite{dietz2025principles} that compromise LLM-as-a-judge approaches.

\paragraph{Future Work}
The ability to accurately encode human rater preferences about texts has benefits beyond being able to judge the relevance of an unjudged document in an information retrieval system without the involvement of the original human judge. Firstly, it enables the assessment of relevance from many more perspectives. We are able to very precisely model a human rater with relatively few judgments from them. Given that in many search scenarios dimensions like trustworthiness and understandability are key to evaluating effectiveness, we plan to investigate if the approach we developed in this paper can be used to model dimensions beyond topical relevance. Another major open problem facing the information retrieval community at the moment is the reliable offline evaluation of RAG responses, as these are generated at query time and thus not able to be evaluated within the Cranfield paradigm. Disregarding the evaluation of attribution and focusing exclusively on the textual content, we plan to investigate the same technique presented in this paper to evaluate the quality of RAG responses according to different utility dimensions.

\raggedright\balance\printbibliography
\end{document}